\documentclass[aps,prl,floatfix,superscriptaddress,reprint]{revtex4-2}

\usepackage{graphicx}
\usepackage{color}
\usepackage{amsfonts}
\usepackage{amssymb}
\usepackage{amsmath}
\usepackage{hyperref}

\definecolor{green}{rgb}{0, 1, 0}

\let\baraccent=\= 
\renewcommand{\=}[1]{\stackrel{#1}{=}} 
\newcommand{\mc}[1]{\mathcal{ #1}} 

\begin{document}

\title{Type-II topological phase transitions of topological skyrmion phases}

\author{Reyhan Ay}
\affiliation{Max Planck Institute for Chemical Physics of Solids, N{\"o}thnitzer Stra{\ss}e 40, 01187 Dresden, Germany}
\affiliation{Max Planck Institute for the Physics of Complex Systems, N{\"o}thnitzer Stra{\ss}e 38, 01187 Dresden, Germany}
\affiliation{Izmir Institute of Technology, Gülbahçe Kampüsü, 35430 Urla Izmir, Türkiye}

\author{Joe H. Winter}
\affiliation{Max Planck Institute for Chemical Physics of Solids, N{\"o}thnitzer Stra{\ss}e 40, 01187 Dresden, Germany}
\affiliation{Max Planck Institute for the Physics of Complex Systems, N{\"o}thnitzer Stra{\ss}e 38, 01187 Dresden, Germany}
\affiliation{SUPA, School of Physics and Astronomy, University of St.\ Andrews, North Haugh, St.\ Andrews KY16 9SS, UK}

\author{A. M. Cook}
\affiliation{Max Planck Institute for Chemical Physics of Solids, N{\"o}thnitzer Stra{\ss}e 40, 01187 Dresden, Germany}
\affiliation{Max Planck Institute for the Physics of Complex Systems, N{\"o}thnitzer Stra{\ss}e 38, 01187 Dresden, Germany}

\begin{abstract}
We present minimal toy models for topological skyrmion phases of matter, which generically realize type-II topological phase transitions in effectively non-interacting systems, those which occur without closing of the minimum direct bulk energy gap. We study the bulk-boundary correspondence in detail to show that a non-trivial skyrmion number yields a rich bulk-boundary correspondence. We observe gapless edge states, which are robust against disorder, due to non-trivial skyrmion number.  Edge states corresponds to bands, which do not traverse the bulk gap, instead yielding gaplessness due to their overlap in energy and exponential localization on opposite edges of the system. These gapless boundary modes can occur for total Chern number zero, and furthermore correspond to rich real-space spin textures with strong polarization of spin along the real-space edge. By introducing toy models generically exhibiting type-II topological phase transitions and characterizing the bulk-boundary correspondence due to non-trivial skyrmion number in these models, we lay the groundwork for understanding consequences of the quantum skyrmion Hall effect.
\end{abstract}

\maketitle

The quantum Hall effect (QHE), in which a two-dimensional electron gas subjected to an out-of-plane external magnetic field exhibits Hall conductivity quantized to rational numbers in units of $e^2/h$~\cite{klitzing1980, PhysRevLett.48.1559}, serves as the foundation for much of our understanding of topological phases of matter~\cite{PhysRevLett.50.1395, FQH-Laughlin, kallin1984excitations, PhysRevLett.62.82, halperin1993theory, halperin1982quantized, halperin1984statistics, wen1991gapless, PhysRevLett.63.199, BERNEVIG2002185, nayak2008, sarma_majorana_2015, freedman_two-eigenvalue_2002, Kitaev_2001}. In particular, it is the foundation for those topological states defined by mappings  to the space of projectors onto occupied states as in the ten-fold way classification scheme~\cite{ryu2010, schnyder2008, PhysRevB.78.195424, Kitaev_2009, altland1997, chiu2016, 3DTI-exp1, chern-discovery, Ersatz-fermi-SPT}. Recently-introduced topological skyrmion phases (TSPs) of matter~\cite{cook2023}, however, instead derive from a much larger set of mappings to the space of myriad observable expectation values, $ \langle \mathcal{O} \rangle$. Such mappings can also be divided into topologically-distinct sectors and understood as lattice counterparts of a generalization of the QHE to the quantum skyrmion Hall effect (QSkHE)~\cite{qskhe}. Ultimately, this relates to the generalization of the concept of a particle: in the QSkHE, point charges of the QHE generalize to point-like quantum skyrmions forming in myriad observable fields, with corresponding generalizations of gauge fields, and $p$-dimensional charges are more generally topological textures in underlying fields. This physics serves as the link between topological states descending from the QHE, and extended magnetic skyrmions~\cite{bogdanov1989thermodynamically, yu_real-space_2010, doi:10.1126/science.1166767, schulz_emergent_2012, doi:10.1126/science.1145799, sampaio_nucleation_2013, donnelly_time-resolved_2020, zhang2020skyrmion, maccariello_electrical_2018, jiang_direct_2017, PhysRevLett.107.136804, PhysRevLett.102.186602, bruno2004topological, PhysRevLett.102.186601}. 

\begin{figure}
   \centering
   \includegraphics[width=\linewidth]{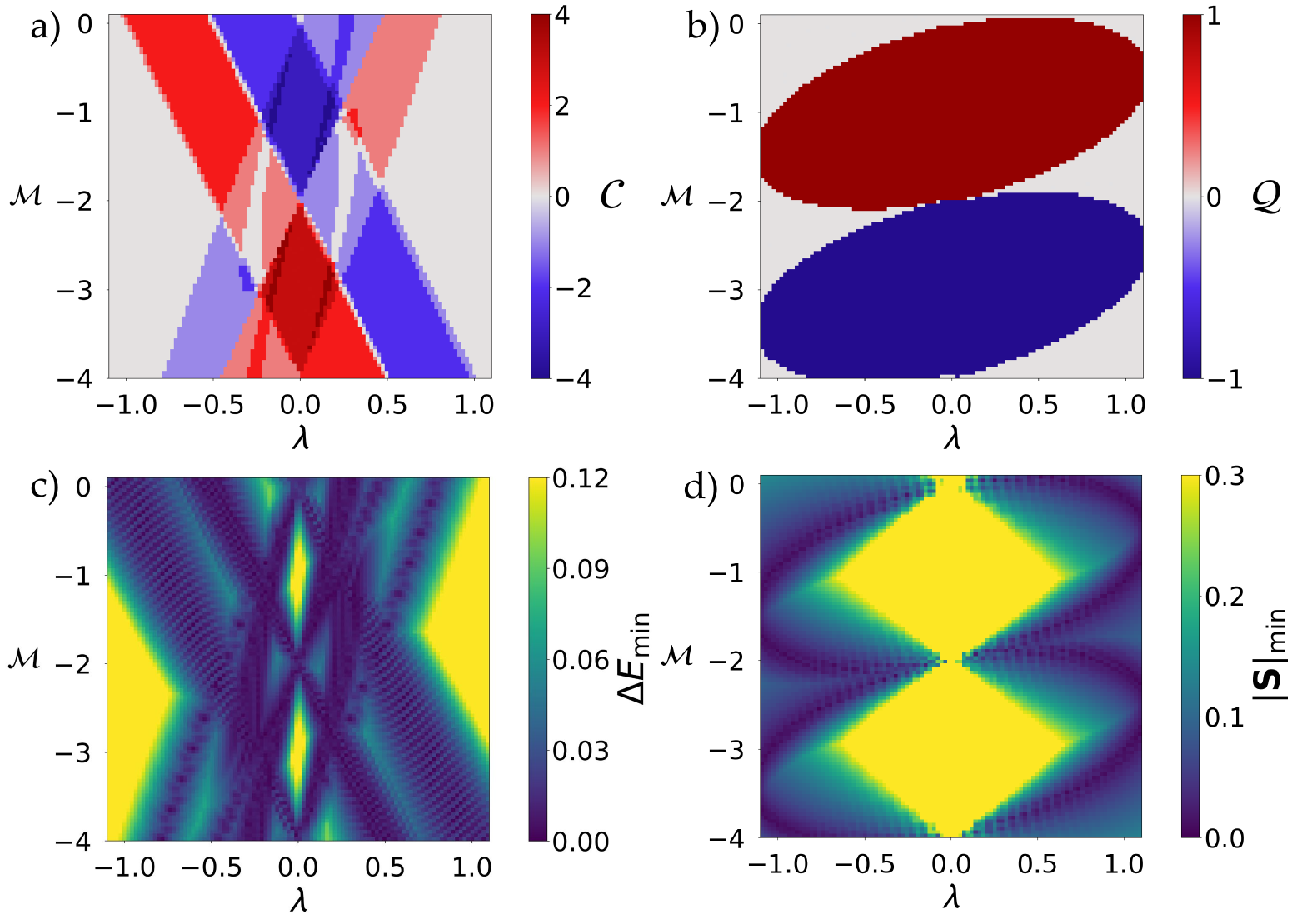}
    \caption{Phase diagrams characterizing bulk topology of Hamiltonian Eq.~\ref{SCZhangtoy} for half-filling, including a) total Chern number $\mathcal{C}$, b) skyrmion number $\mathcal{Q}$, c) minimum direct bulk energy gap $\Delta E_{\mathrm{min}}$, and d) minimum ground state spin magnitude $|\boldsymbol{S}|_{\mathrm{min}}$, with each of these four quantities plotted vs. atomic spin-orbit coupling (SOC) strength $\lambda$ and mass term $M$. Other parameters are fixed at $\beta=0.5$ and $\Delta=0.6$. }
    \label{bulkphsidag}
\end{figure}

A variety of TSPs have been identified~\cite{cook2023,qskhe, liu2023, calderon2023_TRIskyrm}, but the significance of these implications strongly motivates efforts to better understand this physics. A fundamental signature of TSPs is the type-II topological phase transition~\cite{cook2023}, in which a topological invariant changes in effectively non-interacting systems, while maintaining fixed occupancy of the bands and while respecting the symmetries protecting the topological phase, \textit{without the closing of the minimum direct bulk energy gap}. This occurs, for instance, when topological skyrmion phases are realized for mappings to the spin expectation value of occupied states and spin is not conserved due to non-negligible atomic spin-orbit coupling~\cite{cook2023}. In this work, we therefore make a key contribution to understanding this broader set of topological phases associated with the QSkHE, by investigating the type-II topological phase transition in minimal models. We show that type-II topological phase transitions are generic to these models, and also characterize a bulk-boundary correspondence specific to topological skyrmion phases, distinguished by topologically-robust, gapless boundary modes with distinctive spin textures.

\textit{Minimal models for type-II topological phase transitions}---We construct toy models generically realizing type-II topological phase transitions inspired very directly by tight-binding Bogoliubov de Gennes Hamiltonians for superconducting Sr\textsubscript{2}RuO\textsubscript{4} in which topological skyrmion phases were first discovered~\cite{cook2023, ueno2013, Ng_2000}, which we also provide in the Supplementary Materials (Supp. Mat.), Section I, ``Tight-binding model for mirror subsectors of superconducting Sr\textsubscript{2}RuO\textsubscript{4} Bogoliubov de Gennes Hamiltonian and associated spin operators''. Our results are therefore very relevant to transition metal oxide superconductors.

We  consider minimal Bloch Hamiltonians realizing topological skyrmion phases of the following form, with three essential terms,
\begin{equation}
    \mathcal{H}(\boldsymbol{k}) = \boldsymbol{d}(\boldsymbol{k}) \cdot \boldsymbol{\tilde{S}} + \mathcal{H}_{\mathrm{pair}}(\boldsymbol{k}) + \mathcal{H}_{\mathrm{SOC}},
    \label{SCZhangtoy}
\end{equation}
and basis vector
\begin{equation}
    \Psi^{\dagger}_{\boldsymbol{k}} = \left(c^{}_{\boldsymbol{k}, xy, \uparrow}, c^{}_{\boldsymbol{k}, yz, \downarrow} , c^{}_{\boldsymbol{k}, xz, \downarrow}, c^{\dagger}_{\boldsymbol{-k}, xy, \uparrow}, c^{\dagger}_{\boldsymbol{-k}, yz, \downarrow} , c^{\dagger}_{\boldsymbol{-k}, xz, \downarrow}\right),
\end{equation}
where $c_{\boldsymbol{k}, \ell, \sigma}$ annihilates a fermion with momentum $\boldsymbol{k}$ in orbital $\ell$ with spin $\sigma$, and $\{xy, yz, xz\}$ correspond to the t\textsubscript{2g} orbital degree of freedom (dof) and $\{\uparrow, \downarrow\}$ correspond to a spin $1/2$ dof. We note that the unconventional ordering of annihilation and creation operators results from rotating the basis of the Sr\textsubscript{2}RuO\textsubscript{4} tight-binding model to that for which the mirror operator taking $z\rightarrow -z$ is diagonal~\cite{ueno2013, cook2023}.
Here, $\boldsymbol{d}(\boldsymbol{k}) \cdot \boldsymbol{\tilde{S}}$ is the term required to produce a non-trivial skyrmionic texture in the ground state spin expectation value, where $\boldsymbol{\tilde{S}} = \mathrm{diag}\left(\boldsymbol{S}, -\boldsymbol{S}^* \right)$, and $\boldsymbol{S}$ is the vector of spin operators, $\mathcal{S}_x$, $\mathcal{S}_y$, and $\mathcal{S}_z$, for the particle sector. This normal state spin representation is provided in the Supp. Mat., Section I, ``Tight-binding model for mirror subsector I of superconducting Sr\textsubscript{2}RuO\textsubscript{4} Bogoliubov de Gennes Hamiltonian''. The second term, $\mathcal{H}_{\mathrm{pair}}(\boldsymbol{k})$, is a pairing term required to couple the generalized particle-hole conjugate sectors (for a Bogoliubov de Gennes Bloch Hamiltonian, $\mathcal{H}_{\mathrm{pair}}(\boldsymbol{k})$ is the superconducting pairing term). Finally, $\mathcal{H}_{\mathrm{SOC}}$ is the atomic spin-orbit coupling (SOC) term. We find that the first two terms yield only the type-I topological phase transition, and the final SOC term is required to produce the type-II transition as expected from past work.

We focus here on study of six-band models, for which the type-II topological phase transition has already been observed in work introducing the topological skyrmion phases, for a tight-binding model previously used to study Sr\textsubscript{2}RuO\textsubscript{4} with spin-triplet pairing~\cite{Ng_2000, ueno2013, cook2023}. To construct a concrete toy model, we use the spin operators, atomic spin-orbit coupling term, and pairing term of mirror subsector I of the previously-studied Bogoliubov de Gennes Bloch Hamiltonian for Sr\textsubscript{2}RuO\textsubscript{4} with spin-triplet pairing. The form of this mirror subsector Bloch Hamiltonian is reviewed in the Supp. Mat., Section I, ``Tight-binding model for mirror subsector I of superconducting Sr\textsubscript{2}RuO\textsubscript{4} Bogoliubov de Gennes Hamiltonian''. 
    The spin-orbit coupling term of the normal state $h_{SOC}$ and the superconducting gap function of the pairing term $\Delta_1(\boldsymbol{k})$ take the following forms, respectively,
    \begin{align}
h_{SOC} = \begin{pmatrix}
0 & -i\lambda & -\lambda \\
i\lambda & 0 & i\lambda  \\
-\lambda & -i\lambda & 0
\end{pmatrix}, \hspace{2mm}
\Delta_1(\boldsymbol{k}) = \mathrm{diag}\left(\delta_{-}, \delta_{+}, \delta_{+} \right),
\end{align}
where $\lambda$ is atomic spin-orbit coupling strength, $\delta_{\mp} = \left(\Delta_0 / 2\right) \left[ i\sin(k_x) \mp \sin(k_y)\right] $, and $\Delta_0$ is superconducting pairing strength.

  We choose a $\boldsymbol{d}(\boldsymbol{k}) = \langle d_x(\boldsymbol{k}), d_y(\boldsymbol{k}), d_z(\boldsymbol{k}) \rangle$ vector previously used to describe one of the simplest two-band models for a Chern insulator on a square lattice, known as the QWZ model~\cite{qi2006}, 
$d_x(\boldsymbol{k}) = \sin(k_x)$, $d_y(\boldsymbol{k}) = \sin(k_y)$, $d_z(\boldsymbol{k}) = \beta (2 + \mc{M} - \cos(k_x) - \cos(k_x))$.

\textit{Bulk topology}---
For the chosen model given by Eq.~\ref{SCZhangtoy}, we first compute bulk phase diagrams as a function of the mass parameter $\mc{M}$ of the QWZ model and the atomic spin-orbit coupling strength $\lambda$ in parallel to earlier work~\cite{cook2023}. We assume half-filling of the six bands and compute the total Chern number $\mc{C}$, the skyrmion number $\mc{Q}$, the minimum direct bulk energy gap $\Delta E_{\mathrm{min}}$, and the minimum spin magnitude over the Brillouin zone, $|\boldsymbol{S}|_{\mathrm{min}}$, as shown in Fig.~\ref{bulkphsidag} a), b), c) and d), respectively. Additional characterization of bulk topology for a second choice of $\boldsymbol{d}(\boldsymbol{k})$, previously used by Sticlet~\emph{et al.}~\cite{Sticlet2012} to realize two-band Chern insulator phases with Chern number $\mathcal{C} = \pm 2$ for the lower band, is detailed in the Supp. Mat., Section II, ``Additional characterization of bulk topology for a second choice of $\boldsymbol{d}(\boldsymbol{k})$ vector''.

The total Chern number $\mc{C}$ phase diagram, Fig.~\ref{bulkphsidag} a), exhibits a rich variety of non-trivial regions, including a central set of four rhombus-like regions formed by overlapping stripe-like regions. The total Chern number is trivial for large $|\lambda|$ and $\mathcal{M}$ near $-2$. In constrast, the skyrmion number $\mc{Q}$ phase diagram, Fig.~\ref{bulkphsidag} b) is simpler, with two non-trivial regions of $\mc{Q}=\pm1$, respectively. Considering these two phase diagrams together, we find a variety of regions with $\mc{C}$ non-zero and $\mc{Q}$ zero \textit{and vice versa}.

Examining the minimum direct bulk energy gap and minimum spin magnitude over the Brillouin zone shown in Fig.~\ref{bulkphsidag} c) and d), respectively, we find a great variety of both type-I ($\Delta E_{\mathrm{min}}$ goes to zero) and type-II ($\Delta E_{\mathrm{min}}$ remains finite while $|\boldsymbol{S}|_{\mathrm{min}}$ goes to zero) topological phase transitions~\cite{cook2023}. We find such variety in values of the bulk topological invariants and in the topological phase transitions is generic for models of the form Eq.~\ref{SCZhangtoy}, making them invaluable for understanding the interplay between the total Chern number $\mc{C}$ and skyrmion number $\mc{Q}$.

\begin{figure}[h!]
   \centering
   \includegraphics[width=\linewidth]{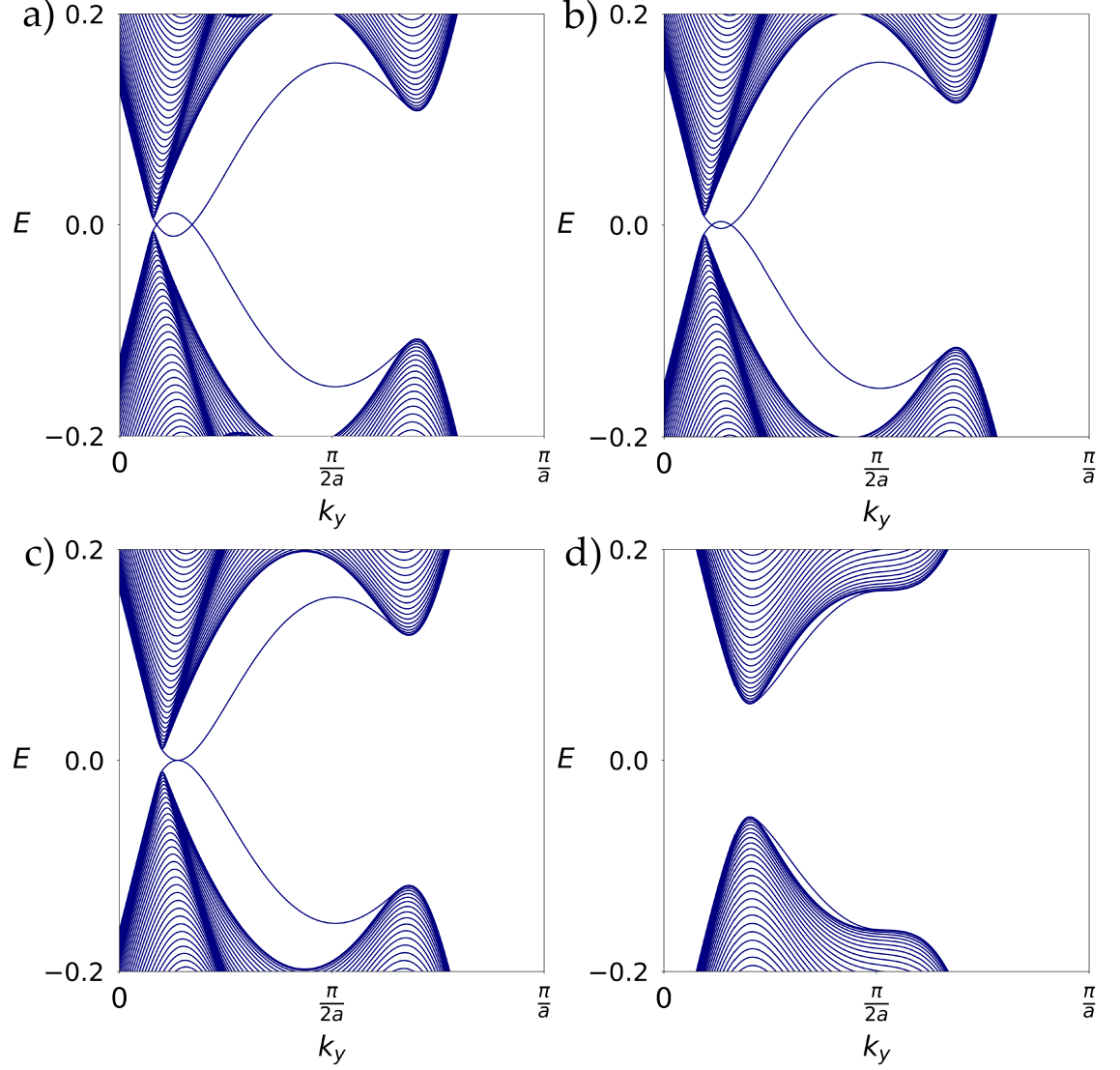}
    
    \caption{Slab spectra for Hamiltonian Eq.~\ref{SCZhangtoy} with open boundary conditions (OBCs) in the $\hat{x}$-direction and periodic boundary conditions (PBCs) in the $\hat{y}$-direction. Overlapping edge states for non-trivial skyrmion number $\mathcal{Q}$ are present as shown in a) and b), which become critical as shown in c) during the type-II topological phase transition, before finally pulling apart fully yielding a gap as shown in d). Parameter values are a) $\mc{M}=-2.7$ and $\lambda=0.8$, b) $\mc{M}=-2.6$ and $\lambda=0.8$, c) $\mc{M}=-2.558$ and $\lambda=0.8$, d) $\mc{M}=-1.9$ and $\lambda=0.8$. }
    
    \label{slabspectracut}
\end{figure}

\textit{Bulk-boundary correspondence}---We now characterize bulk-boundary correspondence for the specific realization of Eq.~\ref{SCZhangtoy} characterized in the bulk in Fig.~\ref{bulkphsidag}, focusing on better understanding the bulk-boundary correspondence associated with changing $\mc{Q}$. To do so, we first consider four points in phase space along a cut through the phase diagram Fig.~\ref{bulkphsidag} labeled `Transition A', for which $\mc{C}=0$ but $\mc{Q}$ changes from -1 to +1, with three other type-II topological phase transitions B, C and D also shown in the Supp. Mat., Section III, ``Additional slab spectra through type-II topological phase transitions''. 

We plot the slab spectrum of the Hamiltonian for open boundary conditions (OBCs) in the $\hat{x}$-direction and periodic boundary conditions (PBCs) in the $\hat{y}$-direction along Transition A in Fig.~\ref{slabspectracut}. Within the $\mc{Q}\neq 0$ region, we observe gaplessness at the edge due to overlapping of two bands in the gap, exponentially-localized on opposite edges. Edge states extend from the bulk valence (conduction) to bulk valence (conduction) bands, similarly to edge states of TSPs in three-band models~\cite{qskhe}. During the type-II topological phase transition, the overlap of these in-gap states shrinks and is finally lost. Gaplessness is lost \textit{in the vicinity of the point} at which $|\boldsymbol{S}_{\mathrm{min}}| = 0$: the separation in phase space between the point at which gaplessness is lost and the point at which $\mc{Q}$ changes in the bulk tends to increase with increasing $|\lambda|$, as shown in the Supp. Mat., Section IV, ``Effect of atomic spin-orbit coupling strength on transition A''. This is a remarkable result in combination with the robustness of these gapless edge states against disorder presented in the next section: although we gain considerable understanding of the spin topology through observable-enriched entanglement~\cite{winter2023_oept} and the view of non-trivial $\mc{Q}$ yielding an effective Chern insulator of the spin sector upon tracing out non-spin degrees of freedom, the entanglement due to non-negligible atomic spin-orbit coupling enriches the physics considerably and requires further investigation. These results may correspond to one of the more general scenarios of the QSkHE, corresponding to topological transport of quantum skyrmion-like textures generalizing the point charges of the QHE, \textit{which are not well-approximated by coarse-graining to point charges}.

While gapless edge states associated with non-trivial $\mc{Q}$ are present in Fig.~\ref{slabspectracut} a) and lost over type-II topological phase transition A as shown in Fig.~\ref{slabspectracut} a) to d), gapless edge states are not observed over this cut through the phase diagram for OBCs instead in the $\hat{y}$-direction and PBCs in the $\hat{x}$-direction. While Transitions B and C are similarly anisotropic, Transition D is more isotropic, with gapless edge states observed for each set of boundary conditions in the vicinity of Transition D. These additional slab spectra are shown in the Supp. Mat., Section III, ``Additional slab spectra through type-II topological phase transitions''. 

\begin{figure}[h!]
   \centering
   \includegraphics[width=\linewidth]{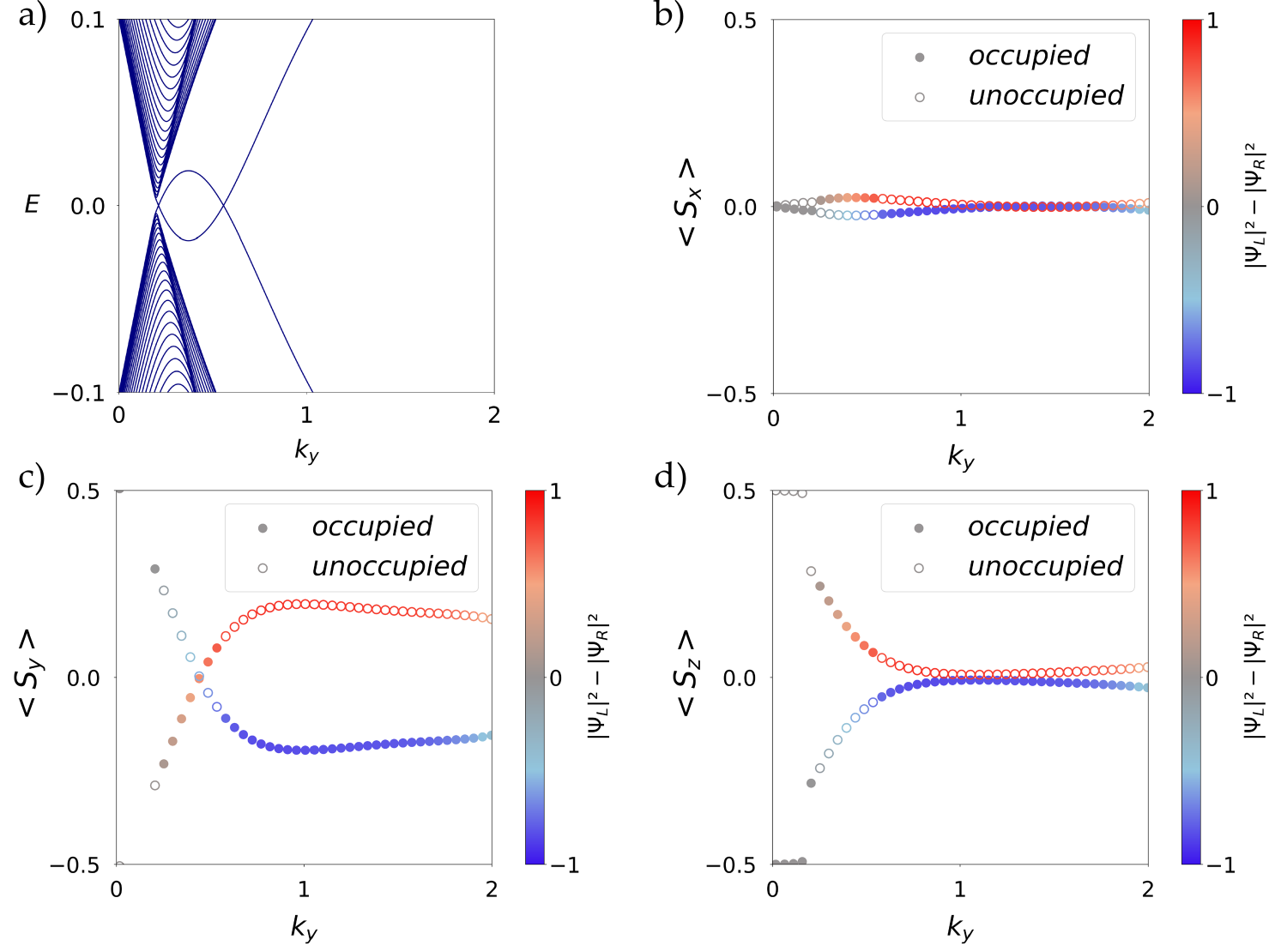}
   
    \caption{ Further characterization of Transition A: a) Slab spectrum for Hamiltonian Eq.~\ref{SCZhangtoy} with parameters $\mc{M} = -2.8$ and $\lambda = 0.8$. $x$-component, $y$-component, and $z$-component of spin expectation value, $\langle S_x(k_y)\rangle$, $\langle S_y(k_y)\rangle$, and $\langle S_z(k_y)\rangle$ for the two overlapping edge states in a) are shown in b), c) and d), respectively. Filled (unfilled) symbols correspond to occupied states at energy $E<0$ (unoccupied states at energy $E>0$). Color bar shows localization of edge states as difference in probability density of a given edge state at the left edge, $|\psi_L|^2$, and right edge, $|\psi_R|^2$.}
    \label{edgespintexture}
\end{figure}

To further characterize the bulk-boundary correspondence, we also compute the spin texture for the gapless edge states, as shown in Fig.~\ref{edgespintexture}. Despite rich physics of the bulk Hamiltonian including non-negligible atomic spin-orbit coupling, the gapless edge states exhibit very strong polarization of spin along the edge. The polarization is chiral, as shown in Fig.~\ref{edgespintexture} c): There is a sign difference in $\langle S_y(k_y) \rangle$ between the two edges. 

Edge state spin textures for each of Transition A to D are shown in the Supp. Mat. Section V, ``Additional edge state spin textures through type-II topological phase transitions''. Edge state spin polarization along sample edges weakens with loss of gaplessness but remains pronounced.

\begin{figure}[h]
   \centering
   \includegraphics[width=\linewidth]{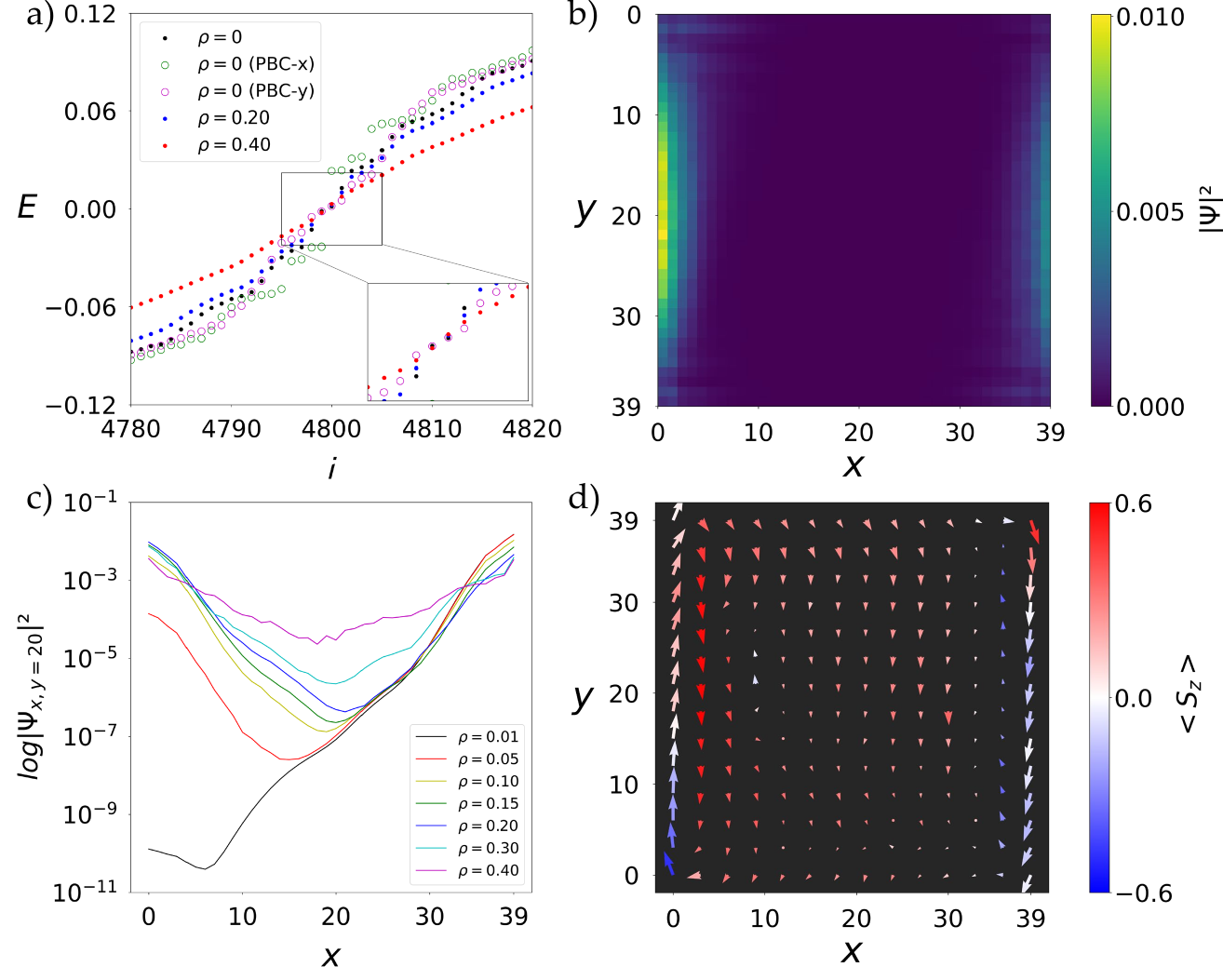}
   
    \caption{a) Spectrum vs. eigenvalue index $i$ for system of size $N_x = N_y = 40$ and parameter set $\mc{M}=-2.6$ and $\lambda=0.8$ (along Transition A cut) for different values of disorder strength $\rho$, for system with OBCs in each of the $\hat{x}$- and $\hat{y}$-directions except for PBC-x (OBC in $\hat{y}$-direction, PBCs in $\hat{x}$-direction), or PBC-y (OBC in $\hat{x}$-direction, PBC in $\hat{y}$-direction). Inset highlights low-energy spectrum. b) Disorder-averaged probability density for eigenstate number i = 4800 from a) for $\rho=0.20$, with 100 disorder realizations. c) Log of disorder-averaged probability density for eigenstate with eigenvalue -0.00121 and i = 4800 vs. position in the $\hat{x}$-direction for $y=20$, shown for different values of $\rho$, with system size and parameter set as in a). d) Disorder-averaged spin texture for eigenstate with eigenvalue -0.00121 {and $i = 4800$} over real-space for $\rho = 0.20$ as in b), with system size and parameter set as in a).}
    
    \label{disorderfig}
\end{figure}

\textit{Robustness of gapless boundary modes against disorder}---We now study the robustness of the gapless edge states due to non-zero $\mc{Q}$ against disorder, by Fourier-transforming the Bloch Hamiltonian and studying the system with OBCs in each of the $\hat{x}$- and $\hat{y}$-directions, along Transition A studied in Fig.~\ref{slabspectracut} and \ref{edgespintexture}. We introduce a spatially-varying on-site potential term that is $\mc{C}'$-invariant at each site $(x,y)$ $\mc{H}_{\mathrm{dis}}(x,y) = \rho \varepsilon(x,y) \mathrm{diag}\left(\mathbf{I}_3, -\mathbf{I}_3 \right)$, where $\rho$ is disorder strength, $\varepsilon(x,y)$ is chosen from a uniform random distribution over the interval $\left[-1,1 \right]$, and $\mathbf{I}_3$ is the $3 \times 3$ identity matrix. The spectrum of the clean system for PBCs and OBCs, as well as the disorder-averaged spectrum for OBCs, is shown in Fig.~\ref{disorderfig} a). Eigenenergy $E$ increases steeply with index $i$ for low-energy states for $\rho=0$, corresponding to edge states, and this slope persists to $\rho=0.2$. At $\rho=0.4$, the slope near $E=0$ is more gradual and bulk-like, indicating collapse of the bulk energy gap. For $\rho=0.2$, we show the disorder-averaged probability distribution over the sample, for the state just below zero in energy of each disorder realization. The state is strongly localized on the $x=0$ and $x=39$ edges.  Fig.~\ref{disorderfig} c) shows a corresponding cut through b) for $y=20$, revealing exponential decay of the probability density into the bulk for $\rho=0.2$, which is lost at $\rho=0.4$, in correspondence with changes observed in Fig~\ref{disorderfig} a). 

We also examine the robustness of edge state spin texture against disorder, as shown in Fig.~\ref{disorderfig} d). We see the strong polarization of the spin along the edge for $\alpha_0=0.2$, but also see a subdominant helicity in the spin components perpendicular to the edge. Additionally, there is some evidence of counter-polarized spin texture just away from the edge. These features are present before disorder-average as shown in the Supp. Mat., Section VI, ``Boundary mode real-space spin textures'', which also, notably, exhibit a real-space, extended \textit{skyrmionic} texture in the bulk. This indicates that the ingredients required to realize topological skyrmion phases, even in relatively simple toy models, also yield extended real-space magnetic skyrmions typically studied in the continuum or in far more complex lattice models~\cite{thore2023}. We defer a more complete study of these rich real-space magnetic textures to a future work.

\textit{Discussion and conclusion}---
We introduce a class of minimal Bloch Hamiltonians generically realizing type-II topological phase transitions of topological skyrmion phases~\cite{cook2023, liu2023, calderon2023_TRIskyrm} and the quantum skyrmion Hall effect~\cite{qskhe}. We also present evidence of gapless edge states due to non-trivial skyrmion number corresponding to bands which do not traverse the bulk gap, and instead yield gaplessness due to their overlap in energy and localization on opposite edges of the system. This gaplessness is topologically-robust against disorder, and corresponds to distinctive edge state spin textures, including a dominant polarization of spin along the edge, and a subdominant feature that appears helical in nature. These arc-like gapless boundary modes also occur generically in other models~\cite{qskhe, winter2023_oept}, indicating results presented here are broadly-relevant to topological skyrmion phases and the quantum skyrmion Hall effect.

\begin{acknowledgments}

\textit{Acknowledgements}---This research was supported in part by the National Science Foundation under Grants No.NSF PHY-1748958 and PHY-2309135, and undertaken in part at Aspen Center for Physics, which is supported by National Science Foundation grant PHY-2210452.

\end{acknowledgments}

\bibliography{p1bib.bib}

\cleardoublepage


\clearpage

\makeatletter
\renewcommand{\theequation}{S\arabic{equation}}
\renewcommand{\thefigure}{S\arabic{figure}}
\renewcommand{\thesection}{S\arabic{section}}
\setcounter{equation}{0}
\setcounter{section}{0}
\onecolumngrid
\begin{center}
  \textbf{\large Supplemental material for ``Type-II topological phase transitions of topological skyrmion phases''}\\[.2cm]
  Reyhan Ay$^{1,2,3}$, Joe H. Winter$^{1,2,4}$, and Ashley M. Cook$^{1,2,*}$\\[.1cm]
  {\itshape ${}^1$Max Planck Institute for Chemical Physics of Solids, Nöthnitzer Strasse 40, 01187 Dresden, Germany\\
  ${}^2$Max Planck Institute for the Physics of Complex Systems, Nöthnitzer Strasse 38, 01187 Dresden, Germany\\
  ${}^3$Izmir Institute of Technology, Gülbahçe Kampüsü, 35430 Urla Izmir, Türkiye\\
  ${}^4$SUPA, School of Physics and Astronomy, University of St.\ Andrews, North Haugh, St.\ Andrews KY16 9SS, UK}
  ${}^*$Electronic address: cooka@pks.mpg.de\\
\end{center}

\section{I. Tight-binding model for mirror subsectors of superconducting Sr\textsubscript{2}RuO\textsubscript{4} Bogoliubov de Gennes Hamiltonian and associated spin operators}\label{Sup_sec_I}

A two-dimensional model for Sr\textsubscript{2}RuO\textsubscript{4}~\cite{ueno2013}  in the $x-y$ plane consists of a Bogoliubov de Gennes Hamiltonian with a spin half degree of freedom, three-fold $t_{2g}$ orbital degree of freedom, and particle-hole degree of freedom, corresponding to $12 \times 12$ matrix representation. This Hamiltonian is invariant under a mirror operation taking $z$ to $-z$, and may be block-diagonalized by going to the basis in which the mirror operator matrix representation is diagonal. We present the Bloch Hamiltonians of these mirror subsectors here, each with $6 \times 6$ matrix representation. We consider the case where each mirror subsector Bloch Hamiltonian itself possesses particle-hole symmetry~\cite{ueno2013,cook2023} corresponding to invariance under charge conjugation $\mc{C}$ ($\mc{C}^2 = -1$), as well as a generalized particle-hole symmetry corresponding to invariance under operation $\mc{C}' = \mc{C}\mc{I}$, where $\mc{I}$ is spatial inversion.

One mirror subsector Bloch Hamiltonian ($6 \times 6$ matrix representation) takes the following form:

\begin{align}
\mc{H}_1(\boldsymbol{k}) = \begin{pmatrix}
h_1(\boldsymbol{k}) & \Delta_1 (\boldsymbol{k}) \\
\Delta_1^{\dagger}  (\boldsymbol{k}) & -h_1^*(-\boldsymbol{k})
\end{pmatrix}.
\end{align}

with basis vector $\Psi^{\dagger}_{\boldsymbol{k}} = \left(c^{}_{\boldsymbol{k}, xy, \uparrow}, c^{}_{\boldsymbol{k}, yz, \downarrow} , c^{}_{\boldsymbol{k}, xz, \downarrow}, c^{\dagger}_{\boldsymbol{-k}, xy, \uparrow}, c^{\dagger}_{\boldsymbol{-k}, yz, \downarrow} , c^{\dagger}_{\boldsymbol{-k}, xz, \downarrow}\right)$ and
\begin{align}
h_1(\boldsymbol{k}) = \begin{pmatrix}
-\mu_B H_z + \mu' + 4 t_3 \cos(k_x)\cos(k_y) + 2 t_2 \left[\cos(k_x) + \cos(k_y)\right] & -i\lambda & -\lambda \\
i\lambda & \mu_B H_z + \mu + 2 t_1 \cos(k_x) & i\lambda + 4 t_4 \sin(k_x)\sin(k_y) \\
-\lambda & -i\lambda + 4 t_4 \sin(k_x)\sin(k_y)  & \mu_B H_z + \mu + 2 t_1 \cos(k_y)
\end{pmatrix}
\end{align}
and
\begin{align}
\Delta_1(\boldsymbol{k}) = \begin{pmatrix}
 {\Delta_0 \over 2} \left[ i\sin(k_x) - \sin(k_y)\right] & 0 & 0 \\
0 & {\Delta_0 \over 2} \left[i \sin(k_x) +\sin(k_y)\right] & 0 \\
0 & 0 & {\Delta_0 \over 2}\left[ i \sin(k_x) + \sin(k_y)\right]
\end{pmatrix}.
\end{align}

Here, $\mu_B H_z$ is strength of applied Zeeman field in the $\hat{z}$-direction, $\mu'$ is an energy offset between $xy$ and $xz$/$yz$ orbitals, $\mu$ is chemical potential, $t_1$, $t_2$, $t_3$, $t_4$ are hopping integrals, and $\lambda$ is the atomic spin-orbit coupling strength.

The basis of this mirror subsector may be used to identify a spin representation provided in past work on topological skyrmion phases~\cite{cook2023}. The spin representation is $\boldsymbol{\tilde{S}} = \mathrm{diag}\left(\boldsymbol{S}, -\boldsymbol{S}^* \right)$, where $\boldsymbol{S}$ is the vector of spin operators, $\mathcal{S}_x$, $\mathcal{S}_y$, and $\mathcal{S}_z$, for the particle sector, taken to be
\begin{align}
    S_x = {1 \over \sqrt{2}}\begin{pmatrix} 0 & 1 & 1 \\
    1 & 0 & 1 \\
    1 & 1 & 0\end{pmatrix}, \hspace{5mm}
     S_y = {1 \over \sqrt{2}}\begin{pmatrix} 0 & -i & -i \\
    i & 0 & -i \\
    i & i & 0\end{pmatrix},  \hspace{5mm}
    S_z = {1 \over \sqrt{2}}\begin{pmatrix} 2 & 0 & 0 \\
    0 & -1 & 0 \\
    0 & 0 & -1\end{pmatrix}.
    \end{align}

For completeness, we also list the second mirror subsector Bloch Hamiltonian although we do not use it as the basis for toy models in this work. This Bloch Hamiltonian takes the form

\begin{align}
\mc{H}_2(\boldsymbol{k}) = \begin{pmatrix}
h_2(\boldsymbol{k}) & \Delta_2 (\boldsymbol{k}) \\
\Delta_2^{\dagger}  (\boldsymbol{k}) & -h_2^*(-\boldsymbol{k})
\end{pmatrix}.
\end{align}

with basis vector $\Phi^{\dagger}_{\boldsymbol{k}} = \left(c^{}_{\boldsymbol{k}, xy, \downarrow}, c^{}_{\boldsymbol{k}, yz, \uparrow} , c^{}_{\boldsymbol{k}, xz, \uparrow}, c^{\dagger}_{\boldsymbol{-k}, xy, \downarrow}, c^{\dagger}_{\boldsymbol{-k}, yz, \uparrow} , c^{\dagger}_{\boldsymbol{-k}, xz, \uparrow}\right)$ and
\begin{align}
h_2(\boldsymbol{k}) = \begin{pmatrix}
\mu_B H_z + \mu' + 4 t_3 \cos(k_x)\cos(k_y) + 2 t_2 \left[\cos(k_x) + \cos(k_y)\right] & -i\lambda & \lambda \\
i\lambda & -\mu_B H_z + \mu + 2 t_1 \cos(k_x) & -i\lambda + 4 t_4 \sin(k_x)\sin(k_y) \\
\lambda & i\lambda + 4 t_4 \sin(k_x)\sin(k_y)  & -\mu_B H_z + \mu + 2 t_1 \cos(k_y)
\end{pmatrix}
\end{align}
and
\begin{align}
\Delta_2(\boldsymbol{k}) = \begin{pmatrix}
 {\Delta_0 \over 2} \left[ i\sin(k_x) + \sin(k_y)\right] & 0 & 0 \\
0 & {\Delta_0 \over 2} \left[i \sin(k_x) -\sin(k_y)\right] & 0 \\
0 & 0 & {\Delta_0 \over 2}\left[ i \sin(k_x) - \sin(k_y)\right]
\end{pmatrix}.
\end{align}

Note that $h_1(-\boldsymbol{k}) = h_1(\boldsymbol{k})$ and $h_2(-\boldsymbol{k}) = h_2(\boldsymbol{k})$. The spin representation for this Hamiltonian is related by symmetry to the representation for $\mc{H}_1(\boldsymbol{k})$ and also provided in earlier work~\cite{cook2023}.

\section{II. Additional characterization of bulk topology for a second choice of $\boldsymbol{d}(\boldsymbol{k})$ vector}\label{Sup_sec_II}
Here, we present a second $\boldsymbol{d}(\boldsymbol{k})$-vector for use in the six-band Bloch Hamiltonian toy model for topological skyrmion phases of matter, and generate phase diagrams using this second $\boldsymbol{d}(\boldsymbol{k})$-vector similar to those shown in Fig.~1 in the main text. We consider the $\boldsymbol{d}(\boldsymbol{k})$-vector of a well-known two-band Bloch Hamiltonian for a Chern insulator, which realizes non-trivial Chern number of $\pm 2$ for the two-band model~\cite{Sticlet2012},
\begin{equation}
    \boldsymbol{d}_s(\boldsymbol{k})= \langle d_x(\boldsymbol{k}), d_y(\boldsymbol{k}), d_z(\boldsymbol{k}) \rangle,
    \label{sticlettoy}
\end{equation} and
\begin{align}
    d_x(\boldsymbol{k}) &= \alpha\cos(k_x), \\ \nonumber
    d_y(\boldsymbol{k}) &= \alpha\cos(k_y), \\ \nonumber
    d_z(\boldsymbol{k}) &= \beta\cos(k_x+k_y)
    \label{sticlettoydxdydz}
\end{align} 

\begin{figure}[h!]
   \centering
   \includegraphics[width=0.8\linewidth]{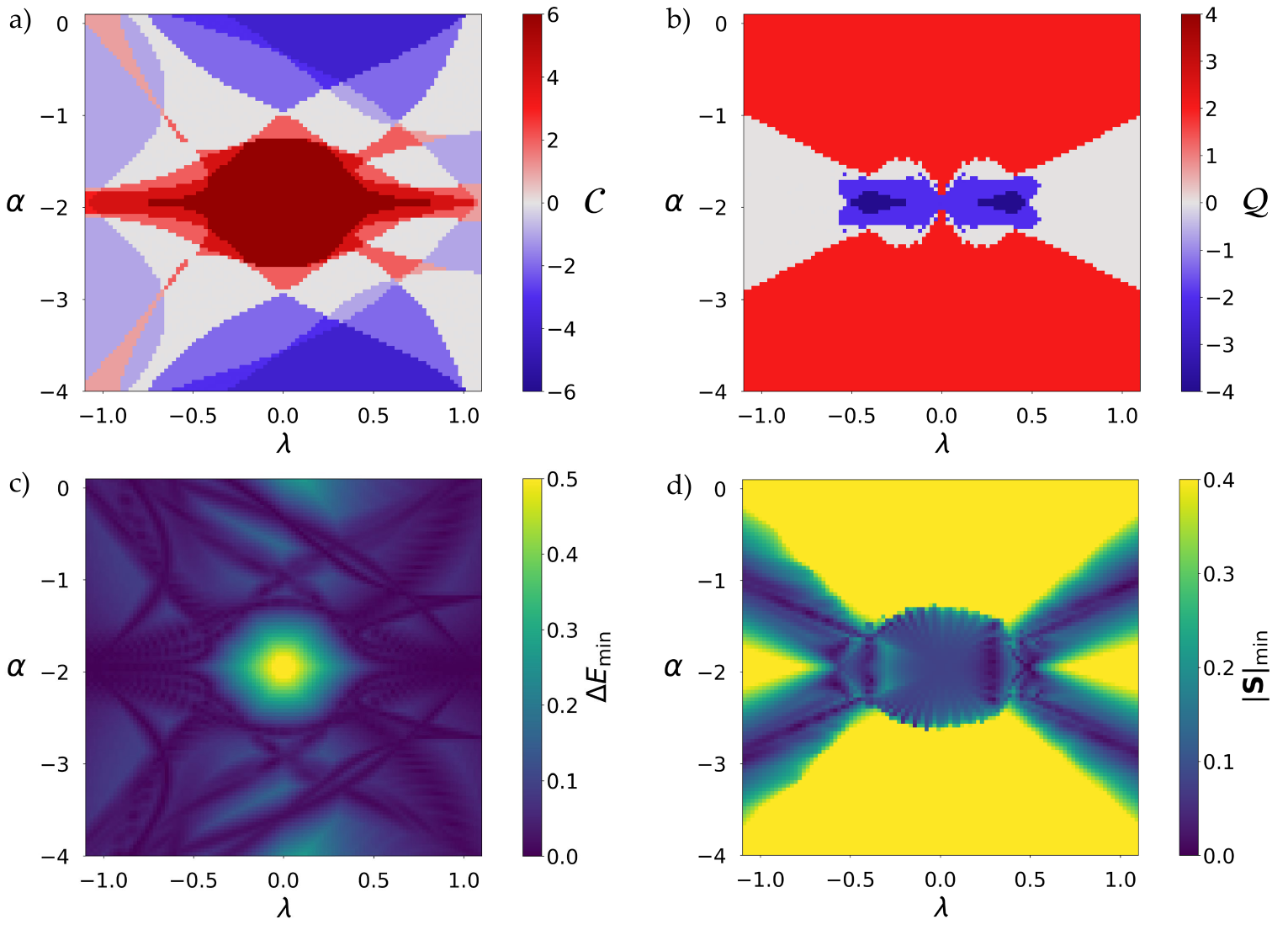}
    \caption{Phase diagrams characterizing bulk topology of Hamiltonian  Eq.~\ref{SCZhangtoy} with $\boldsymbol{d}(\boldsymbol{k})$-vector Eq.~\ref{sticlettoy} for half-filling, including a) total Chern number $\mathcal{C}$, b) skyrmion number $\mathcal{Q}$, c) minimum direct bulk energy gap $\Delta E_{\mathrm{min}}$, and d) minimum ground state spin magnitude $|\boldsymbol{S}|_{\mathrm{min}}$, with each of these four quantities plotted vs. atomic spin-orbit coupling (SOC) strength $\lambda$ and hopping parameter $\alpha$. Other parameters are fixed at $\Delta=0.6$, and $\beta=1$} 
    \label{sticlet}
\end{figure}

\newpage

\section{III. Additional slab spectra through type-II topological phase transitions}\label{Sup_sec_III}

Here, we present additional results on evolution of the spectrum and bulk-boundary correspondence through type-II topological phase transitions for open boundary conditions (OBC) in the $\hat{x}$- ($\hat{y}$)-direction and periodic boundary conditions (PBC) in the $\hat{y}$- ($\hat{x}$)-direction. Transition A for OBC in the $\hat{x}$-direction is shown in the main text, Fig.~2, so here we show additional results in this case for Transitions B to D in Figs.~\ref{slabspectra_Binx}, ~\ref{slabspectra_Cinx}, \ref{slabspectra_Dinx}, respectively. We then show evolution of slab spectra for OBC in the $\hat{y}$-direction for all Transitions A, B, C, and D in Figs.~\ref{slabspectra_Ainy}, ~\ref{slabspectra_Biny}, \ref{slabspectra_Ciny}, and \ref{slabspectra_Diny}, respectively.

\subsection{Transitions B to D, OBC in x}

\begin{figure}[h!]
   \centering
   \includegraphics[width=1\linewidth]{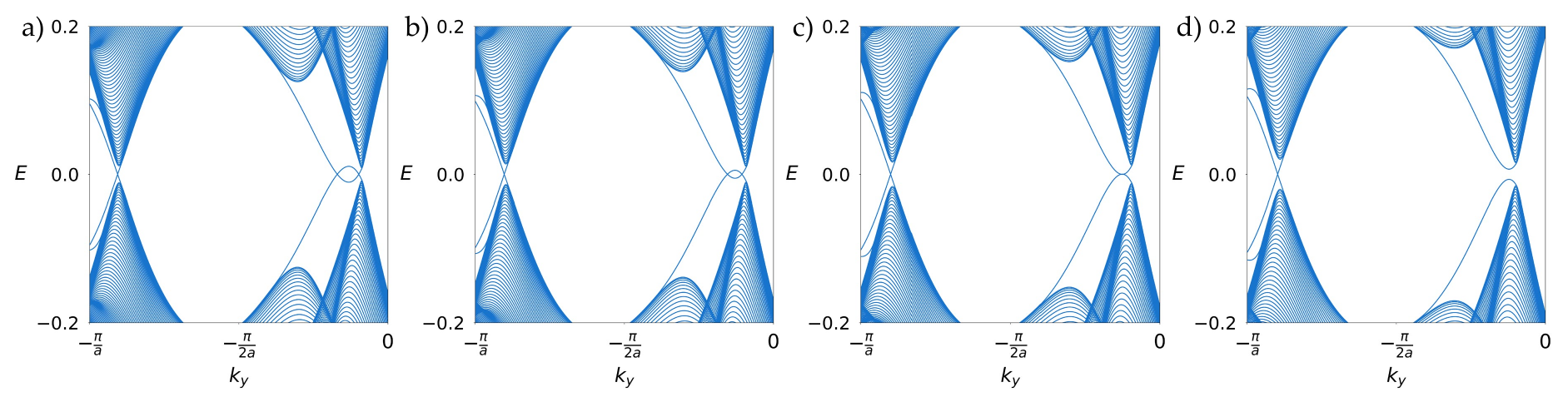}
    \caption{Slab spectra for open boundary conditions in the $\hat{x}$-direction, periodic boundary conditions in the $\hat{y}$-direction across  Transition B for parameters a) $\mc{M}=0.3625$ and $\lambda=-0.75$, b) $\mc{M}=0.125$ and $\lambda=-0.7$, c) $\mc{M}=-0.08875$ and $\lambda=-0.655$, d) $\mc{M}=-0.35$ and $\lambda=-0.60$.}
    \label{slabspectra_Binx}
\end{figure}

\begin{figure}[h!]
   \centering
   \includegraphics[width=1\linewidth]{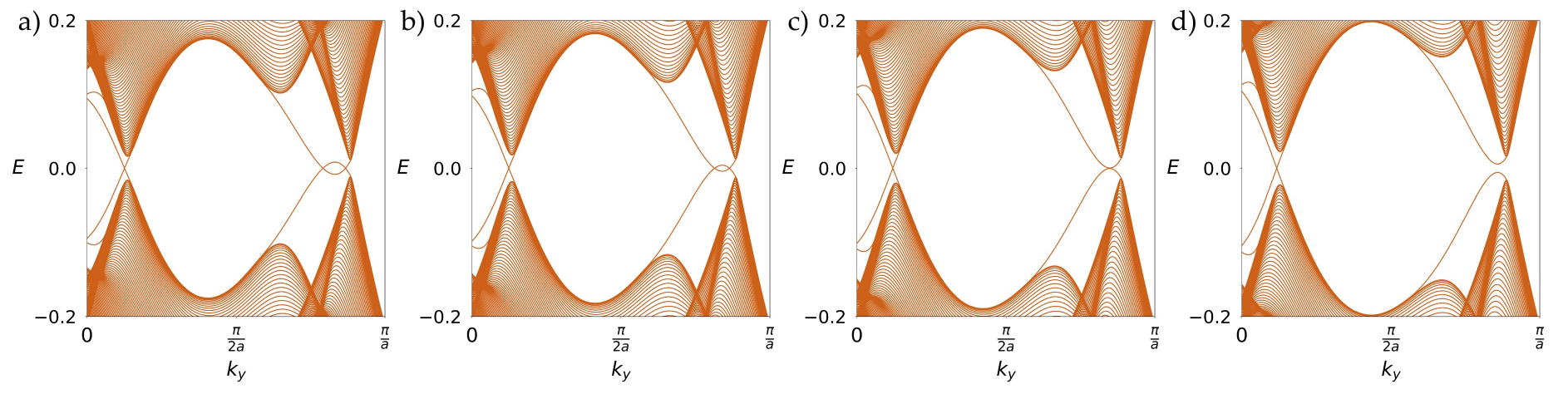}
    \caption{Slab spectra for open boundary conditions in the $\hat{x}$-direction, periodic boundary conditions in the $\hat{y}$-direction across Transition C for parameters a) $\mc{M}=-5.10$ and $\lambda=0.9875$, b) $\mc{M}=-4.7$ and $\lambda=0.8875$, c) $\mc{M}=-4.35$ and $\lambda=0.8$, d) $\mc{M}=-4.0$ and $\lambda=0.7125$.}
    \label{slabspectra_Cinx}
\end{figure}

\begin{figure}[h!]
   \centering
   \includegraphics[width=1\linewidth]{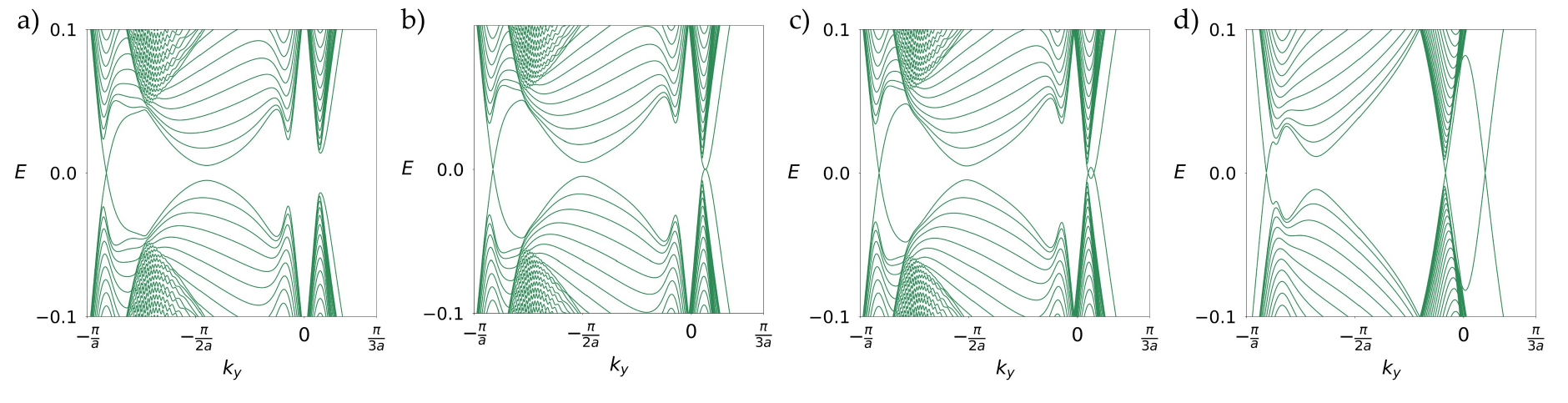}
    \caption{Slab spectra for open boundary conditions in the $\hat{x}$-direction, periodic boundary conditions in the $\hat{y}$-direction across Transition D for parameters a) $\mc{M}=-0.9$ and $\lambda=0.35$, b) $\mc{M}=-1.06$ and $\lambda=0.35$, c) $\mc{M}=-1.10$ and $\lambda=0.35$, d) $\mc{M}=-1.90$ and $\lambda=0.35$.}
    \label{slabspectra_Dinx}
\end{figure}

Here, we show evolution of slab spectra for OBC in the $\hat{x}$-direction for Transitions B, C, and D in  \ref{slabspectra_Binx}, \ref{slabspectra_Cinx}, and \ref{slabspectra_Dinx}, respectively.

\newpage
\subsection{Transitions A to D, OBC in y}

\begin{figure}[h!]
   \centering
   \includegraphics[width=1\linewidth]{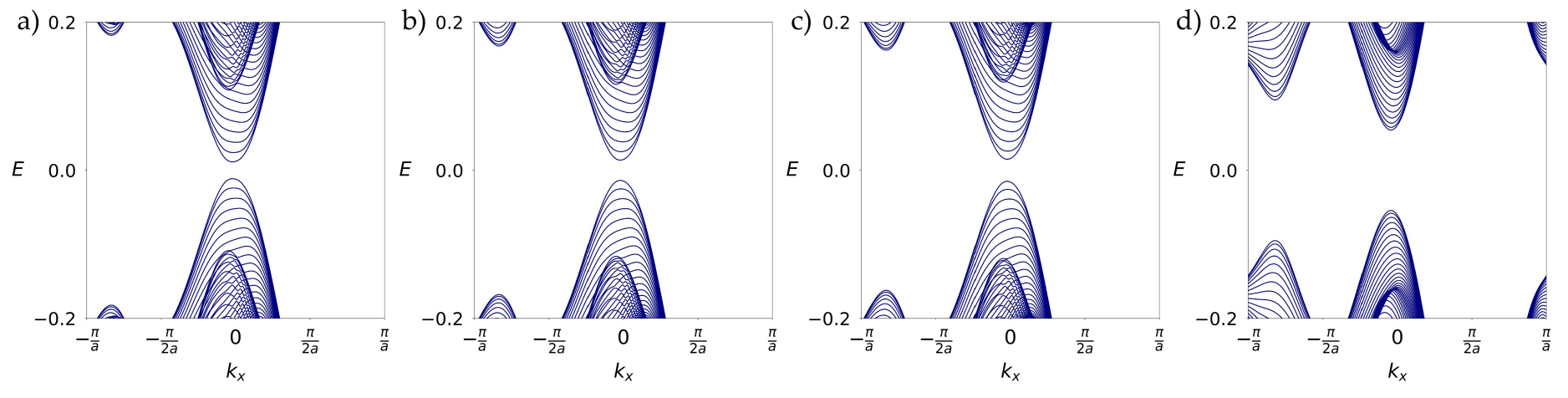}
    \caption{Slab spectra for open boundary conditions in the $\hat{y}$-direction, periodic boundary conditions in the $\hat{x}$-direction across Transition A for parameters a) $\mc{M}=-2.7$ and $\lambda=0.8$, b) $\mc{M}=-2.6$ and $\lambda=0.8$, c) $\mc{M}=-2.558$ and $\lambda=0.8$, d) $\mc{M}=-1.9$ and $\lambda=0.8$.}
    \label{slabspectra_Ainy}
\end{figure}

\begin{figure}[h!]
   \centering
   \includegraphics[width=1\linewidth]{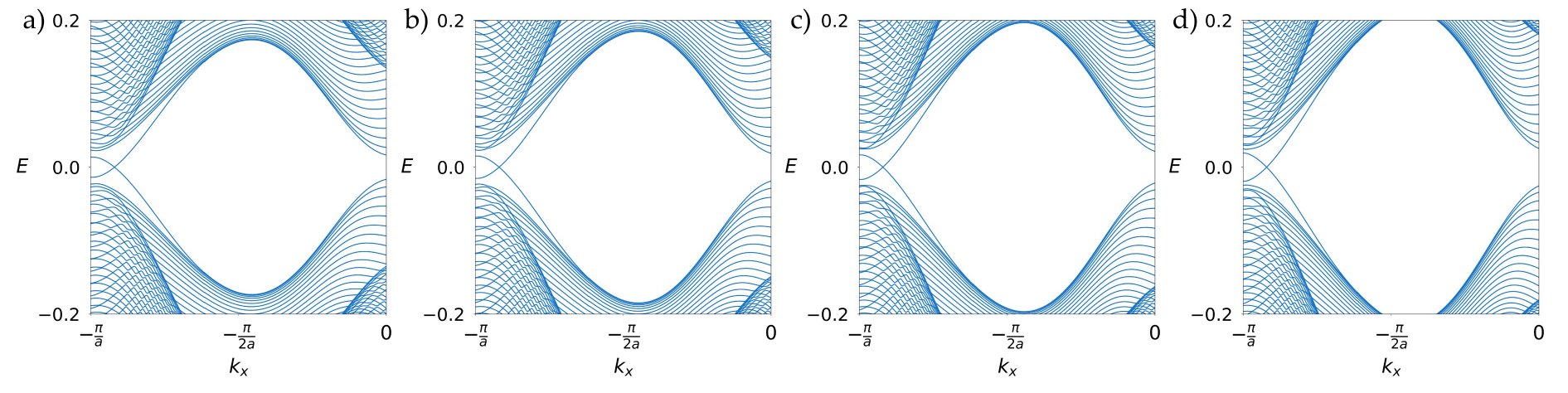}
    \caption{Slab spectra for open boundary conditions in the $\hat{y}$-direction, periodic boundary conditions in the $\hat{x}$-direction across Transition B for parameters a) $\mc{M}=0.3625$ and $\lambda=-0.75$, b) $\mc{M}=0.125$ and $\lambda=-0.7$, c) $\mc{M}=-0.08875$ and $\lambda=-0.655$, d) $\mc{M}=-0.35$ and $\lambda=-0.60$.}
    \label{slabspectra_Biny}
\end{figure}

\begin{figure}[h!]
   \centering
   \includegraphics[width=1\linewidth]{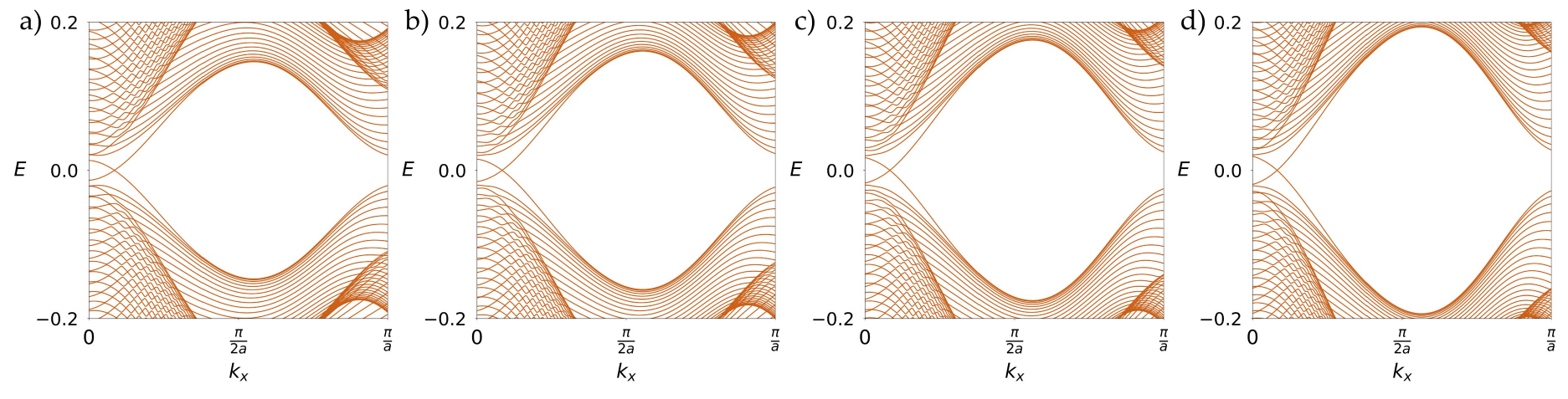}
    \caption{Slab spectra for open boundary conditions in the $\hat{y}$-direction, periodic boundary conditions in the $\hat{x}$-direction across Transition C for parameters a) $\mc{M}=-5.10$ and $\lambda=0.9875$, b) $\mc{M}=-4.7$ and $\lambda=0.8875$, c) $\mc{M}=-4.35$ and $\lambda=0.8$, d) $\mc{M}=-4.0$ and $\lambda=0.7125$.}
    \label{slabspectra_Ciny}
\end{figure}

\begin{figure}[h!]
   \centering
   \includegraphics[width=1\linewidth]{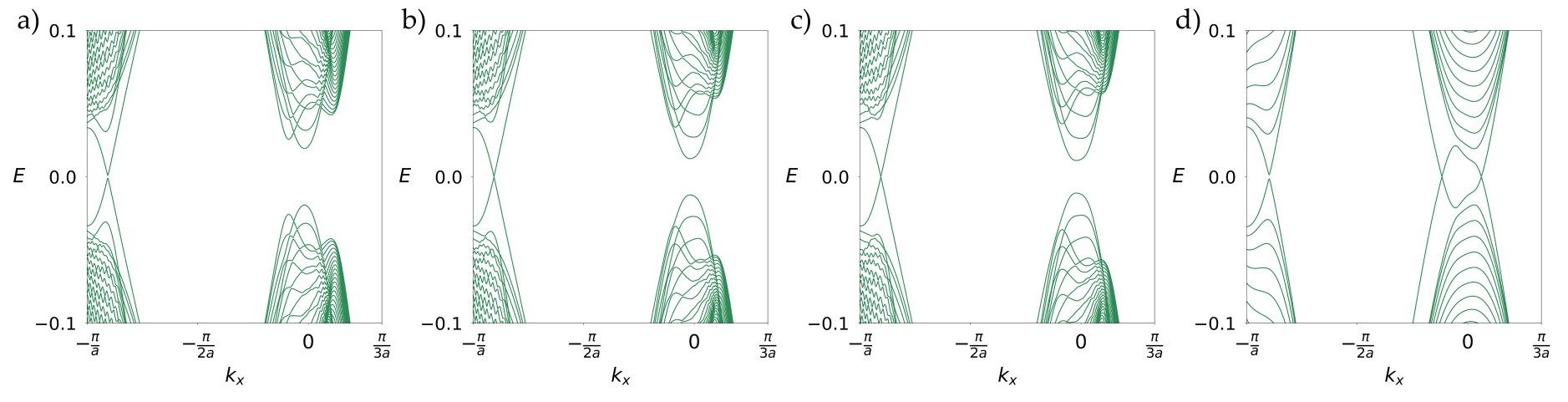}
    \caption{Slab spectra for open boundary conditions in the $\hat{y}$-direction, periodic boundary conditions in the $\hat{x}$-direction across Transition D for parameters a) $\mc{M}=-0.9$ and $\lambda=0.35$, b) $\mc{M}=-1.06$ and $\lambda=0.35$, c) $\mc{M}=-1.10$ and $\lambda=0.35$, d) $\mc{M}=-1.90$ and $\lambda=0.35$.}
    \label{slabspectra_Diny}
\end{figure}

Here, we show evolution of slab spectra for OBC in the $\hat{y}$-direction for all Transitions A, B, C, and D in Figs.~\ref{slabspectra_Ainy}, \ref{slabspectra_Biny}, \ref{slabspectra_Ciny}, and \ref{slabspectra_Diny}, respectively. As shown in  Figs.~\ref{slabspectra_Ainy}, \ref{slabspectra_Biny}, and \ref{slabspectra_Ciny}, Transitions A-C slab spectra for these boundary conditions are qualitatively unchanged across the type-II transition for parameter sets over which topological-robust gaplessness is gained/lost for OBCs instead in the $\hat{x}$-direction, while the slab spectrum for Transition D exhibits additional band-crossings at zero-energy across the type-II transition also for these boundary conditions as shown in  Fig.~\ref{slabspectra_Diny}.

\subsection{Additional results for Transition D with open boundary conditions in $\hat{y}$-direction}

\begin{figure}[h!]
   \centering
   \includegraphics[width=0.75\linewidth]{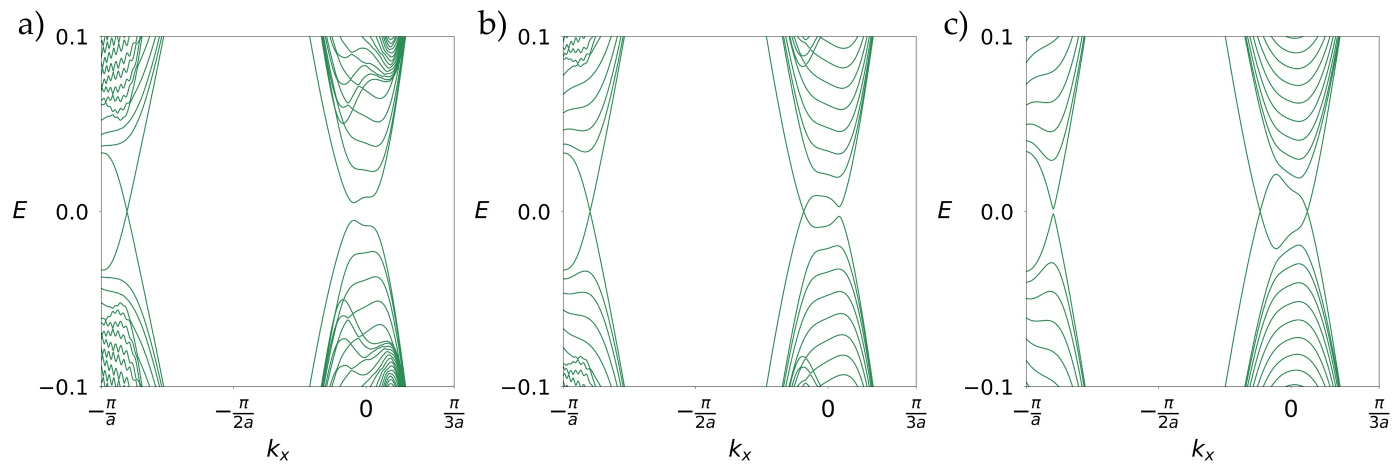}
    \caption{Slab spectra for open boundary conditions in the $\hat{y}$-direction, periodic boundary conditions in the $\hat{x}$-direction highlighting edge state band connectivity across Transition D for parameters a) $\mc{M}=-1.30$ and $\lambda=0.35$, b) $\mc{M}=-1.60$ and $\lambda=0.35$, c) $\mc{M}=-1.90$ and $\lambda=0.35$}
    \label{xtraslabspec_Diny}
\end{figure}

We show additional results highlighting particular features of the gapless boundary states associated with the type-II topological phase transition in Fig.~\ref{xtraslabspec_Diny}. In Fig.~\ref{xtraslabspec_Diny} a), states are present within the bulk gap in the vicinity of $k_x=0$, but do not overlap to yield gaplessness. With even more negative $\mc{M}$, gaplessness occurs through crossing of these in-gap bands at a single $\boldsymbol{k}$-point as shown in Fig.~\ref{xtraslabspec_Diny} b). At even more negative $\mc{M}$, gaplessness persists via crossing of the in-gap bands around $k_x = 0$ at two separate $\boldsymbol{k}$-points, while gaplessness near $k_x=-\pi/a$ is lost.

\section{IV. Effect of atomic spin-orbit coupling strength on transition A}\label{Sup_sec_IV}

\begin{figure}[h!]
   \centering
   \includegraphics[width=1\linewidth]{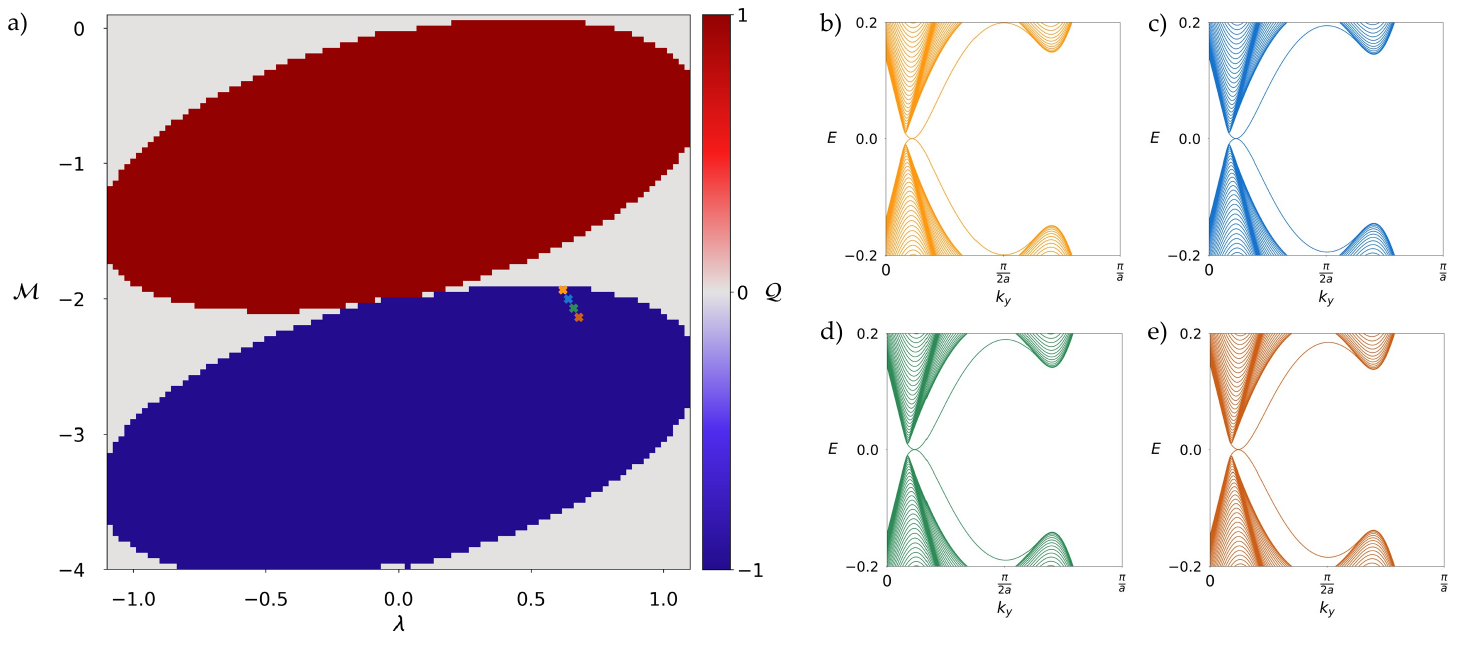}
    \caption{Onset of gaplessness with varying atomic spin-orbit coupling strength. a) depicts the skyrmion number $\mc{Q}$ vs. atomic spin orbit coupling strength $\lambda$ and mass term $\mc{M}$ also shown in Fig.~1 b), with phase space coordinates for four slab spectra computed with open boundary conditions in the $\hat{x}$-direction and periodic boundary conditions in the $\hat{y}$-direction shown in b) for $\mc{M}=-2.136$, $\lambda =0.68$, c) $\mc{M}=-2.069$, $\lambda =0.66$, d) $\mc{M}=-2.001$, $\lambda =0.64$, e) $\mc{M}=-1.934$, $\lambda =0.62$.}
    \label{SOCstrength}
\end{figure}

Here, we show the bulk skyrmion number phase diagram Fig.~\ref{bulkphsidag} b) in Fig.~\ref{SOCstrength} a), highlighting four points in phase space with corresponding slab spectra for OBC in the $\hat{x}$-direction shown in Fig.~\ref{SOCstrength} b), c), d) and e), respectively. Each slab spectrum depicts edge states which just touch rather than overlap, with gaplessness lost through fine-tuned increase in mass parameter $\mc{M}$. In this region, the separation in phase space between when the skyrmion number $\mc{Q}$ changes in the bulk and loss of gaplessness at the edge increases with increasing atomic spin-orbit coupling strength $\lambda$.

\newpage

\section{V. Additional edge state spin textures through type-II topological phase transitions}\label{Sup_sec_V}

\begin{figure}[h!]
   \centering
   \includegraphics[width=1\linewidth]{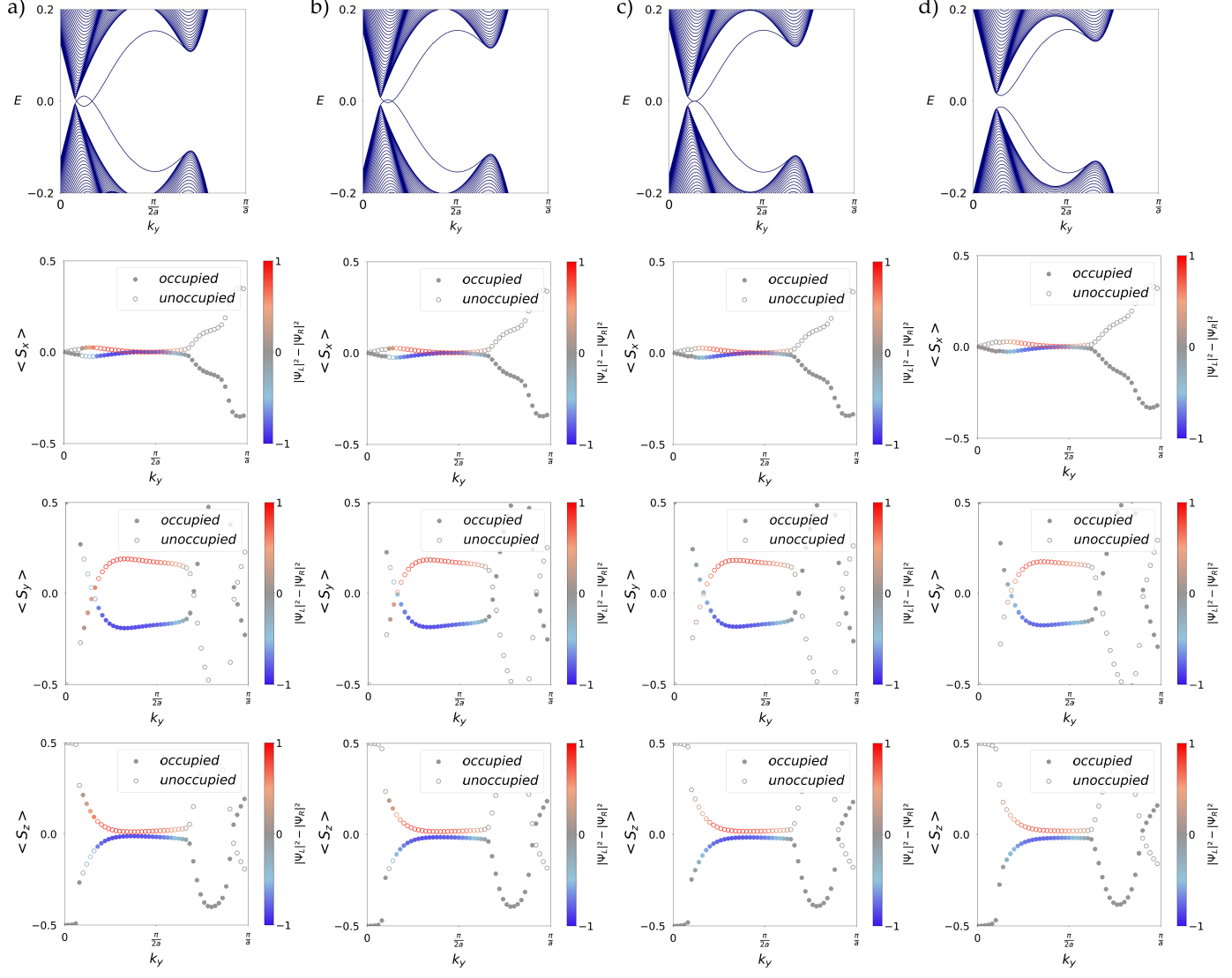}
    \caption{Edge state spin textures, with associated slab spectra, for open boundary conditions in the $\hat{x}$-direction, periodic boundary conditions in the $\hat{y}$-direction, for Transition A. Parameter sets are a) $\mc{M}=-2.7$ and $\lambda=0.8$, b) $\mc{M}=-2.6$ and $\lambda=0.8$, c) $\mc{M}=-2.558$ and $\lambda=0.8$, d) $\mc{M}=-2.4$ and $\lambda=0.8$.}
    \label{transitionA_spintextevo}
\end{figure}

\begin{figure}[h!]
   \centering
   \includegraphics[width=1\linewidth]{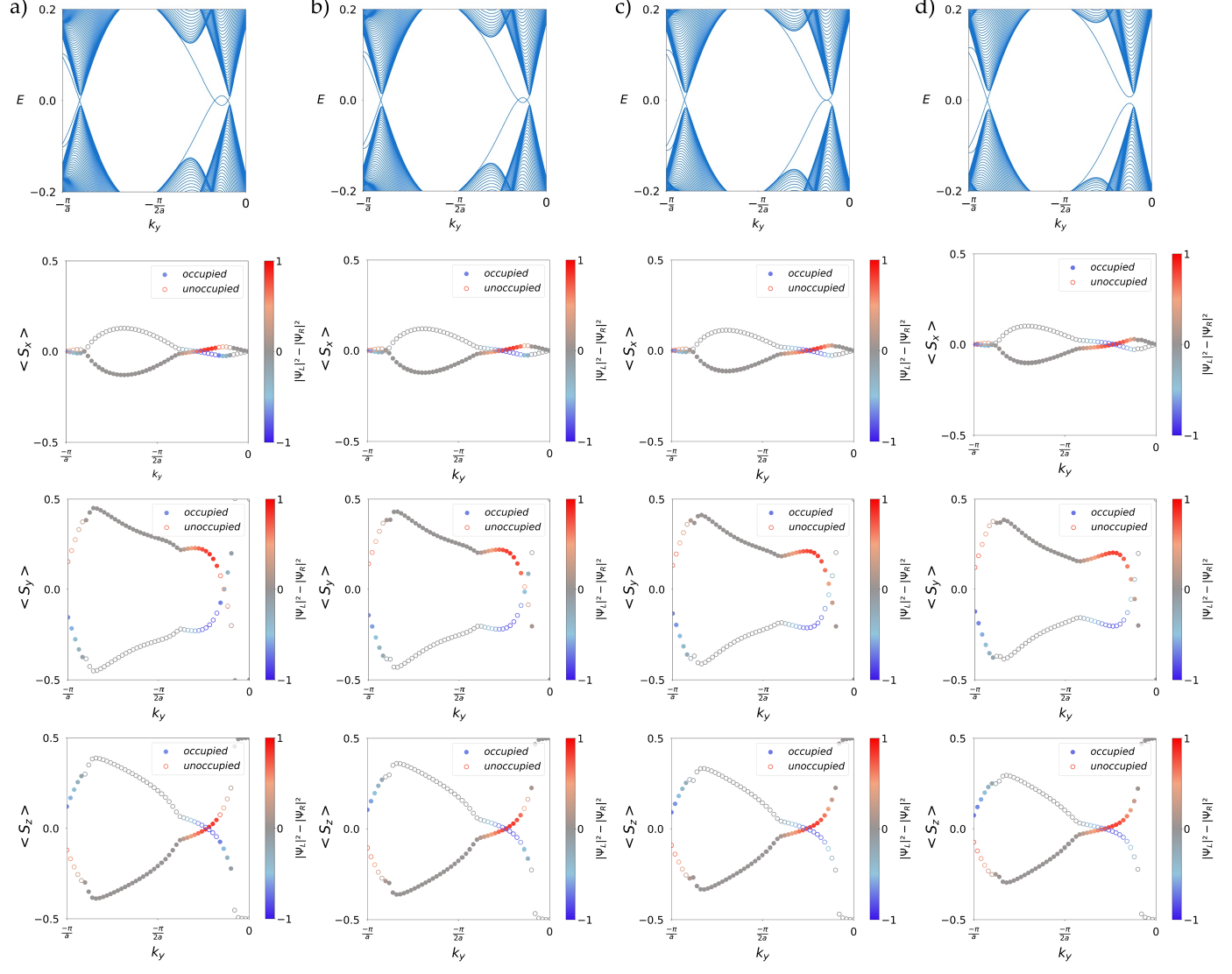}
    \caption{Edge state spin textures, with associated slab spectra, for open boundary conditions in the $\hat{x}$-direction, periodic boundary conditions in the $\hat{y}$-direction, for Transition B. Parameter sets are a) $\mc{M}=0.3625$ and $\lambda=-0.75$, b) $\mc{M}=0.125$ and $\lambda=-0.7$, c) $\mc{M}=-0.08875$ and $\lambda=-0.655$, d) $\mc{M}=-0.35$ and $\lambda=-0.60$.}
    \label{transitionB_spintextevo}
\end{figure}

\begin{figure}[h!]
   \centering
   \includegraphics[width=1\linewidth]{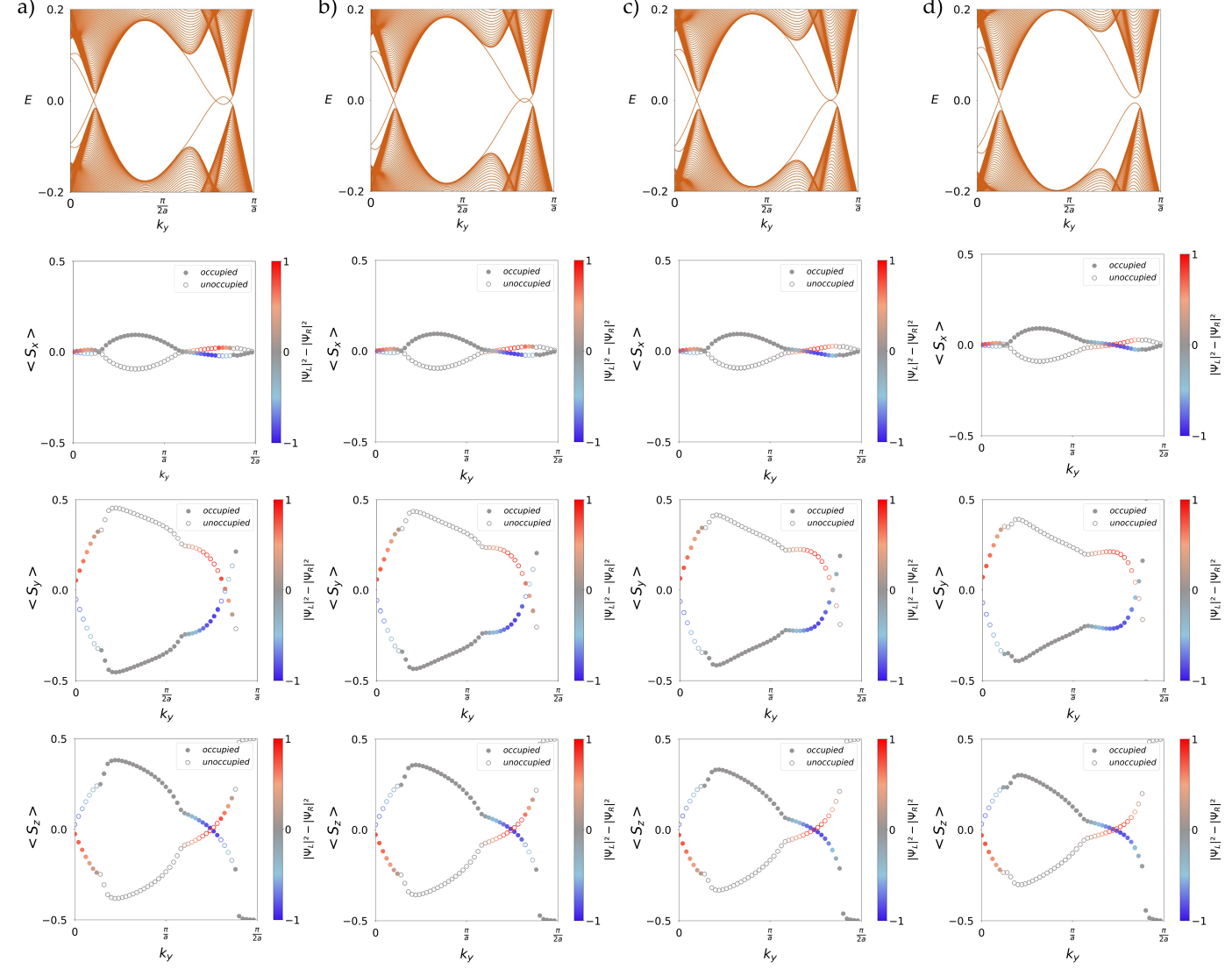}
    \caption{Edge state spin textures, with associated slab spectra, for open boundary conditions in the $\hat{x}$-direction, periodic boundary conditions in the $\hat{y}$-direction, for Transition C. Parameter sets are a) $\mc{M}=-5.10$ and $\lambda=0.9875$, b) $\mc{M}=-4.7$ and $\lambda=0.8875$, c) $\mc{M}=-4.35$ and $\lambda=0.8$, d) $\mc{M}=-4.0$ and $\lambda=0.7125$.}
    \label{transitionC_spintextevo}
\end{figure}

\begin{figure}[h!]
   \centering
   \includegraphics[width=1\linewidth]{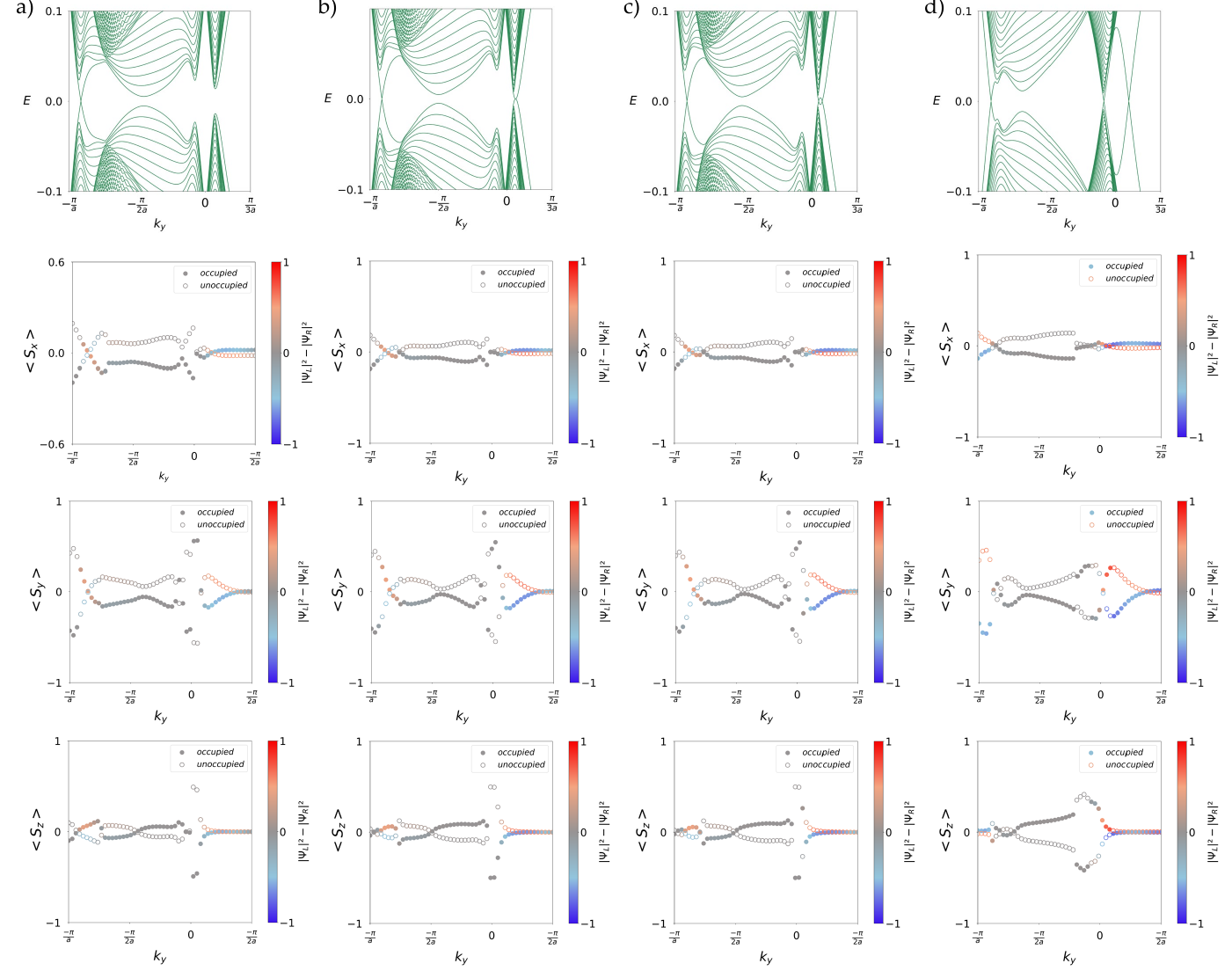}
    \caption{Edge state spin textures, with associated slab spectra, for open boundary conditions in the $\hat{x}$-direction, periodic boundary conditions in the $\hat{y}$-direction, for Transition D. Parameter sets are a) $\mc{M}=-0.9$ and $\lambda=0.35$, b) $\mc{M}=-1.06$ and $\lambda=0.35$, c) $\mc{M}=-1.10$ and $\lambda=0.35$, d) $\mc{M}=-1.90$ and $\lambda=0.35$.}
    \label{transitionD_spintextevo}
\end{figure}

In this section, we show evolution of the edge state spin textures through four type-II topological phase transitions A to D shown in Figs.~\ref{transitionA_spintextevo}, \ref{transitionB_spintextevo}, \ref{transitionC_spintextevo}, \ref{transitionD_spintextevo}, respectively.

\subsection{Transition A}
\newpage

\subsection{Transition B}
\newpage

\subsection{Transition C}

\newpage
\subsection{Transition D}

\newpage

\section{VI. Boundary mode real-space spin textures}\label{Sup_sec_VI}

Here, we show real-space spin textures for individual eigenstates of Hamiltonian Eq.~\ref{SCZhangtoy} without disorder ($\rho=0$) for OBCs in each of the $\hat{x}$- and $\hat{y}$-directions in correspondence with Fig.~\ref{disorderfig} in the main text.
\begin{figure}[h!]
   \centering
   \includegraphics[width=1\linewidth]{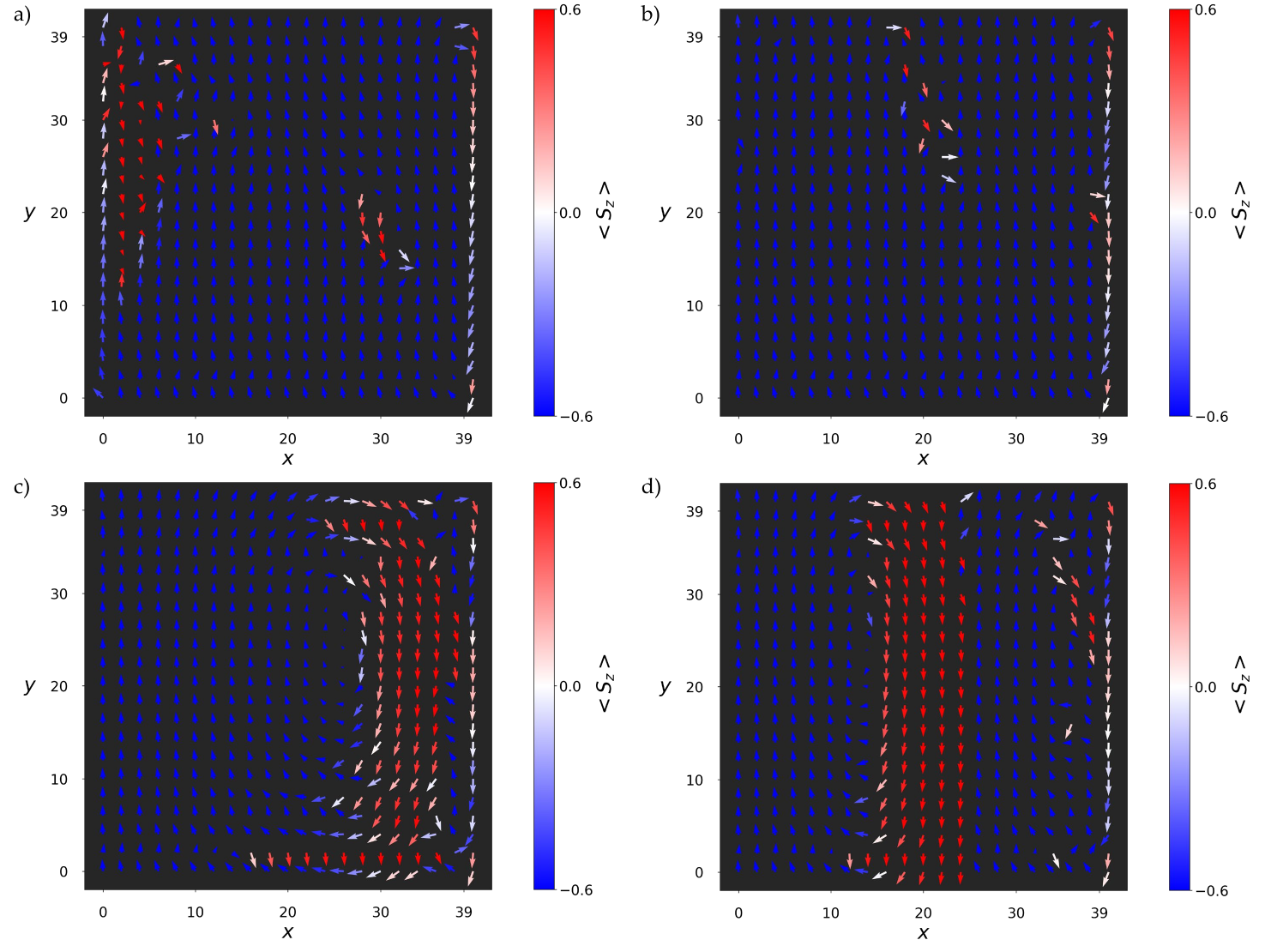}
    \caption{Real-space spin textures of eigenstates number a) i = 4800 at eigenvalue -0.00121, b) i = 4799 at eigenvalue -0.01217, c) i = 4798 at eigenvalue -0.02288, and d) i = 4797 at eigenvalue -0.02486.}
\end{figure}

\end{document}